\documentclass[aps,pra,superscriptaddress,showkeys]{revtex4}

\def\half{\frac{1}{2}}

\usepackage{graphicx}
\usepackage{dcolumn}
\usepackage{bm}
\usepackage{amsthm}
\usepackage{epsfig}

\def\tr{{\rm tr}\, }
\newcommand{\begth}{\begin{theorem}}
\newcommand{\enth}{\end{theorem}}
\newcommand{\bpr}{\begin{proof}}
\newcommand{\epr}{\end{proof}}
\newcommand{\be}{\begin{equation}}
\newcommand{\ee}{\end{equation}}
\newcommand{\bes}{\begin{equation*}}
\newcommand{\ees}{\end{equation*}}
\newcommand{\bea}{\begin{eqnarray}}
\newcommand{\eea}{\end{eqnarray}}
\newcommand{\beas}{\begin{eqnarray*}}
\newcommand{\eeas}{\end{eqnarray*}}

\newtheorem{theorem}{Theorem}
\begin{document}


\title{Calculating a maximizer for
quantum mutual information\\}

\author{Tony Dorlas}
\email{dorlas@stp.dias.ie}
\author{Ciara Morgan}
\email{cqtciara@nus.edu.sg}
\affiliation{%
School of Theoretical Physics\\
Dublin Institute for Advanced Studies\\10 Burlington Road\\ Dublin
4, Ireland.
}%

\date{January 31, 2008}

\begin{abstract}
We obtain a maximizer for the quantum mutual information for
classical information sent over the quantum qubit amplitude damping
channel. This is achieved by limiting the ensemble of input states
to antipodal states, in the calculation of the product-state
capacity for the channel, the resulting maximizing ensemble consisting
of just two non-orthogonal states. We also consider the product-state
capacity of a convex combination of two memoryless channels and
demonstrate in particular that it is in general not given by the
minimum of the capacities of the respective memoryless channels.
\end{abstract}


\maketitle

\section{Introduction}
In this paper we obtain the product-state capacity of the amplitude
damping channel. It is determined by a transcendental equation in
a single real variable, which is easily solved numerically. We
also consider a convex combination of two memoryless channels and
show in particular that the product-state capacity of a convex
combination of a depolarizing and an amplitude damping channel,
which was shown in \cite{1} to be given by the supremum of
the minimum of the corresponding Holevo quantities, is not equal
to the minimum of their product-state capacities.

\subsection{Memoryless channels and the HSW theorem}
The transmission of classical information over a quantum channel
is achieved by encoding the information as quantum states. A
memoryless channel is given by a completely positive
trace-preserving map ${\Phi}: {\cal S}({\cal H}) \to {\cal
S}({\cal K})$, where ${\cal S}({\cal H})$ and ${\cal S}({\cal K})$
denote the states on the input and output Hilbert spaces ${\cal
H}$ and $\cal K$ respectively. In the case of product-state
inputs, the HSW theorem, proved independently by Holevo \cite{2}
and by Schumacher and Westmoreland \cite{3}, states that the
\emph{product-state capacity} for classical information sent
through a memoryless quantum channel is given by
\begin{equation}\label{HSW}
\chi^*(\Phi) = \max_{\{p_j,\rho_{j}\}}
\chi(\Phi)(\{p_j,\rho_j\}),
\end{equation}
where the Holevo-$\chi$-quantity is defined by \bea\label{holevo}
\chi(\Phi)(\{p_j,\rho_j\}) \mathrel{\mathop:}= S\left(\sum_j p_j\, {\Phi}(\rho_{j})\right) - \sum_j p_j \, S\left({\Phi}(\rho_{j})\right), \eea  and where $S$ is the von Neumann
entropy, $S(\rho) = -\tr \left( \rho \, log \, \rho\right)$. The
maximum is taken over all ensembles of input states $\rho_j$ with
probabilities $p_j$.
The capacity for channels with entangled input
states has been studied \cite{4}, and it has been shown that for
certain channels the use of entangled states can enhance the
inference of the output state and increase the capacity (e.g. \cite{5}). We concentrate here on the \emph{product-state} capacity for noisy quantum channels.

Note that, by concavity of the entropy, the maximum in Equation (\ref{HSW})
is always attained for an ensemble of \emph{pure states} $\rho_j$.
Moreover, it follows from Carath\'eodory's theorem (see
\cite{6,7,8}), that the
ensemble can always be assumed to contain no more than $d^2$ pure
states, where $d ={\rm dim}\,({\cal H})$.

\begin{figure}[ht]
\centerline{ \epsfig{file=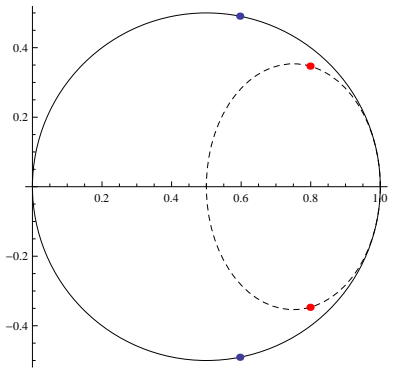, scale=.5}}
\caption{Optimal input-states (blue) to the amplitude-damping channel with $\gamma = 0.5$ and the resulting output states from the channel (red).}\label{ellPlot}
\end{figure}

In Section \ref{substitution} we show that, in the case of the amplitude damping
channel, the maximum is in fact obtained for an ensemble of two pure states
\cite{footNote}. Moreover, these states are in general not orthogonal as in the channel
considered by Fuchs \cite{10}. Figure \ref{ellPlot} demonstrates the action of the amplitude damping channel with error
parameter $\gamma = \half$ with the optimal input-states represented in blue and the corresponding output states in red.

\subsection{Convex combination of memoryless channels}

In \cite{1} the product-state capacity of a
convex combination of memoryless channels was determined. Given a
finite collection of memoryless channels $\Phi_1,
\dots,\Phi_M$ with common input Hilbert space $\cal H$ and
output Hilbert space $\cal K$, a convex combination of these
channels is defined by the map \be \Phi^{(n)}\left(
\rho^{(n)}\right)  = \sum_{i=1}^M \gamma_i
\,\Phi_i^{\otimes n}(\rho^{(n)}), \ee where $\gamma_i,\
(i=1,\dots,M)$ is a probability distribution over the channels
$\Phi_1,\dots,\Phi_M$. Thus, a given input state
$\rho^{(n)} \in {\cal S}({\cal H}^{\otimes n})$ is sent down one
of the memoryless channels with probability $\gamma_i$. This
introduces long-term memory, and as a result the capacity of the
channel $\Phi^{(n)}$ is no longer given by the maximum of
the Holevo quantity. Instead, it was proved in \cite{1}
that it is given by \be C_p(\Phi^{(n)}) =
\sup_{\{p_j,\rho_{j}\}} \left[ \bigwedge_{i=1}^M
\chi_i(\{p_j,\rho_j\}) \right], \label{convexcap} \ee where
$\chi_i= \chi({\Phi}_i)$ is the Holevo quantity for the
$i$-th channel $\Phi_i$.

\section{The amplitude-damping channel and the Holevo-$\chi$-quantity.}\label{substitution}

The qubit amplitude-damping channel models the loss of energy in a qubit
quantum system and is described, with error parameter $0 \leq \gamma \leq 1$, by the following
operation elements
\be E_0 = \left(
\begin{array}{cc}
1 & 0  \\
0 & \sqrt{1 -\gamma}   \\
\end{array} \right), \; \;
E_1 = \left( \begin{array}{cc}
0 & \sqrt{\gamma}  \\
0 & 0   \\
\end{array} \right).\ee
Using the operation elements above, the qubit amplitude-damping channel can be expressed as follows
$\Phi_{amp}(\rho) = E_0 \, \rho
\, E_0^* + E_1 \, \rho \, E_1^*$.
Note that since $E^*_0 E_0 + E^*_1 E_1 = I$, the operator $\Phi_{amp}$ is a CPT map and therefore a legitimate quantum channel.

Acting on the general qubit state $ \rho = \left(
\begin{array}{cc} a & b \\ {\bar b} & 1-a \end{array} \right) \label{gen_state}$,
the amplitude-damping channel $\Phi_{amp}$ is given by
\be
\Phi_{amp}(\rho) = \left(\begin{array}{cc}
a + (1-a) \gamma & b\sqrt{1 - \gamma}  \\
\bar{b}\sqrt{1- \gamma} & (1-a)(1- \gamma)   \\
\end{array} \right).\ee
The eigenvalues of $\Phi_{amp}(\rho)$ are easily found to be
\bea\label{amp_eig} \lambda_{amp\pm} &=& \frac{1}{2} \left(1 \pm
\sqrt{\left(1+2a(\gamma -1) -2\gamma \right)^2 -4|b|^2 (\gamma -1)}
\right). \eea
To maximize the Holevo quantity, given by Equation (\ref{holevo}), for this
channel we show that the first term is increased, while keeping
the second term fixed, if each pure state $\rho_j$ is replaced by
itself and its mirror image in the real $b$-axis, i.e. if we
replace $ \rho_j = \left(
\begin{array}{cc}
a_j & b_j  \\
\bar{b}_j & (1-a_j)   \\
\end{array} \right)$ associated with probability $p_j$, with the states
$\rho_j = \left( \begin{array}{cc}
a_j & b_j  \\
\bar{b_j} & (1-a_j)   \\
\end{array} \right)$ and $\rho_j' = \left(\begin{array}{cc}
a_j & -b_j  \\
-\bar{b}_j & (1-a_j)  \\
\end{array} \right) $, both with probabilities $p_j/2$.

In general, the states $\rho_j$ must lie inside the Poincar\'e
sphere $ \left(a-\half\right)^2 + |b|^2 \leq \frac{1}{4} $ and so
the pure states will lie on the boundary $ |b|^2 = a(1-a)$.

We first show that the second term in Equation (\ref{holevo}) remains
unchanged when the states are replaced in the way described above.
Indeed, since the eigenvalues (\ref{amp_eig}) depend only on
$|b|$, we have $S\left(\Phi(\rho_j)\right) =
S\left(\Phi(\rho_j')\right)$ and therefore the first term
is unchanged. Secondly, by concavity and the fact that $ S \left(
\sum_j p_j \, \Phi(\rho_j') \right) = S \left( \sum_j p_j \,
\Phi(\rho_j) \right)$, we get,
\be
S\left(\sum_j \frac{p_j}{2} \,
\Phi(\rho_j + \rho_j') \right) \geq  S \left( \Phi
\left( \sum_j p_j \, \rho_j \right) \right).
\ee We can conclude that
the first term in Equation (\ref{holevo}) is increased with the second
term fixed if each state $\rho_j$ is replaced by itself together
with its mirror image.

\subsection{Convexity of the output entropy}

We concentrate here on proving that, in the case of the amplitude-damping channel, the second term in the equation for the
Holevo-$\chi$-quantity is convex as a function of the parameters
$a_j$ when $\rho_j$ is taken to be a pure state, i.e. $b_j =
\sqrt{a_j(1-a_j)}$. (Note that $S(a)$ only depends on $|b|$.) Thus
$S\left( \Phi(\rho_j)\right)$ is a function of one variable
only, i.e. $S(a_j) = S (\Phi_{amp}(\rho_{a_j}))$, with
$\rho_a = \left(\begin{array}{cc}
a  & \sqrt{a(1-a)} \\
\sqrt{a(1-a)} & 1-a
\end{array} \right) $ and hence \begin{equation} \sigma(a) =
\Phi_{amp}(\rho_a) = \left(\begin{array}{cc}
a + (1-a) \gamma & \sqrt{a(1-a)}\sqrt{1 - \gamma}  \\
\sqrt{a(1-a)}\sqrt{1- \gamma} & (1-a)(1- \gamma)
\end{array} \right). \label{sigma} \end{equation}
The eigenvalues of (\ref{sigma}) are given by $\lambda_{amp\pm} =
\frac{1}{2} (1 \pm x)$, where $
x=\sqrt{1-4\gamma(1-\gamma)(1-a)^2}$, and thus $ S(a) =
H\left(\frac{1-x}{2}\right)$, where $H(p) = - p \log p - (1-p) \log (1-p)$ is
the binary entropy. It is now easy to see that $S''(a) \geq 0$ and
hence that $S(a)$ is convex. Writing $\bar{\rho_{a}} = \sum_j p_j\,
\rho_{a_j}$ with ${\bar a} = \sum_j p_j\, a_j$ and $\chi_{AD}(\{p_j,\rho_j\}) = \chi(\Phi_{amp})(\{p_j,\rho_j\})$ we have
\begin{equation} \chi_{AD}(\{p_j,\rho_j\})
= S(\Phi_{amp}({\bar{\rho_{a}}})) - \sum_j p_j\,S(a_j) \leq
S(\Phi_{amp}({\bar{\rho_{a}}})) - S({\bar a}). \end{equation}
The capacity is therefore given by \be \chi(\Phi_{amp}) =
\max_{a \in [0,1]} \left[ S \left( \frac{1}{2} (\sigma(a) +
\sigma'(a)) \right) - S(\sigma(a)) \right]. \ee
The maximizing value of $a$ is given by the transcendental equation
$\chi_{AD}'(a)=0$ and can only be computed numerically.

\begin{figure}[h]
          \centerline{
            \epsfig{file=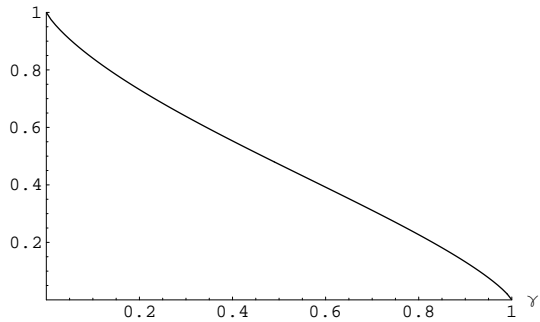, scale=0.7}}
             \caption{The classical capacity of the qubit amplitude
             damping channel plotted as a function of $\gamma$. }
             \label{ADcap}
       \end{figure}

It turns out that $a_{\rm max} \geq \half$ for all $\gamma$. This is in
fact easily proved: The determining equation is \be \chi_{AD}'(a)
\ln 2 = -(1-\gamma) \ln \frac{a + \gamma (1-a)}{(1-\gamma)(1-a)} +
\frac{2 \gamma (1-\gamma) (1-a)}{x} \ln \frac{1+x}{1-x} = 0.
\label{amax} \ee Since $\chi_{AD}(a)$ is concave, the statement
follows if we show that $\chi_{AD}'(\half)
> 0$. But, if $a = \half$, $x = \sqrt{1-\gamma+\gamma^2}$ and
\be
\chi_{AD}'(a=0.5) = - (1-\gamma) \ln \frac{1+\gamma}{1-\gamma} +
\frac{\gamma (1-\gamma)}{x} \ln \frac{1+x}{1-x} > 0
\ee because $x
> \gamma$ and the function $ \frac{1}{2x} \ln \frac{1+x}{1-x} =
\frac{\tanh^{-1}(x)}{x} $ is increasing. The resulting capacity is plotted in Figure \ref{ADcap}.

\section{Convex combinations of two memoryless channels}\label{convexSection}

Let us now consider a convex combination of two memoryless
channels. It was shown in \cite{1} that the product-state
capacity is given by Equation (\ref{convexcap}). Note that we always have
\be
C_p(\Phi^{(n)}) \leq \bigwedge_{i=1}^M \left[ \sup_{\{ p_j, \rho_j\}} \chi_i(\{ p_j, \rho_j\}) \right].
\ee We now
consider three cases: a convex combination of two depolarizing
channels, two amplitude-damping channels, and one depolarizing and
one amplitude-damping channel.

\subsection{Two depolarizing channels}

In the case of a convex combination of two depolarizing qubit
channels $\Delta_{\lambda_i}(\rho) = (1 - \lambda_i) \rho + \lambda_i
(\frac{I}{2})$ with parameters $\lambda_1$ and
$\lambda_2$, we have \be C(\Phi_{\lambda_1,\lambda_2}^{(n)}) =
\chi^*(\lambda_1) \wedge \chi^*(\lambda_2) = \chi^*(\lambda_1 \vee
\lambda_2). \ee Indeed, since the maximizing ensemble for both
channels is the same, namely two projections onto orthogonal
states, this also maximizes the minimum $\chi_1 \wedge \chi_2$.
(The product-state capacity of a depolarizing qubit channel is
well-known of course, and is given by $ \chi^*(\Delta_{\lambda}) = 1 - H
\left(\frac{\lambda}{2} \right)$. In fact, it was proved by King
\cite{11}, that this is also the classical (ultimate) capacity of the channel.)

\subsection{Two amplitude-damping channels}

A convex combination of amplitude-damping channels is similar. In
that case, the maximizing ensemble does depend on the parameter
$\gamma$, but as can be seen from Figure \ref{DepAmpDamp}, for any $a$,
$\chi_{AD}(a)$ decreases with $\gamma$, so $\chi(\gamma_1) \wedge
\chi(\gamma_2) = \chi(\gamma_1 \vee \gamma_2)$ and we have again,
\be C_p(\Phi_{\gamma_1,\gamma_2}^{(n)}) = \chi^*(\gamma_1)
\wedge \chi^*(\gamma_2) = \chi^*(\gamma_1 \vee \gamma_2). \ee In
fact, for $\gamma \leq \half$ this can be seen as follows. The
derivative with respect to $\gamma$ is given by \be \frac{\partial
\chi}{\partial \gamma} = -(1-a) \ln \frac{a + \gamma
(1-a)}{(1-\gamma)(1-a)} + \frac{(2\gamma-1)(1-a)^2}{x} \ln
\frac{1+x}{1-x}. \ee Clearly, if $\frac{a}{1-a} > 1-2\gamma$ both
terms are negative. Otherwise, we remark that $ x \geq
(1-2\gamma)(1-a)$ so that it suffices if $ x > y = 1-2\gamma -
2a(1-\gamma) > 0. $ This is easily checked.

In case $\gamma > \half$, we need to show that $$ f(a,\gamma) =
\ln \frac{a + \gamma(1-a)}{(1-\gamma)(1-a)} -
\frac{(2\gamma-1)(1-a)}{x} \ln \frac{1+x}{1-x} \geq 0. $$ Now, if
$a=0$, then $f(0,\gamma) = 0$, and the derivative is given by
\begin{equation} \frac{\partial f(a,\gamma)}{\partial a} =
\frac{1-\gamma}{a + \gamma(1-a)} + \frac{1}{1-a} + \frac{2
\gamma-1}{x^3} \ln \frac{1+x}{1-x} -
\frac{2(2\gamma-1)}{x^2} \end{equation} which can be shown
to be positive.

\subsection{A depolarizing channel and an amplitude-damping
channel}

We now investigate the product-state capacity of a convex
combination of an amplitude-damping and a depolarizing channel.
Let $\chi_1$ and $\chi_2$ denote the Holevo quantity of the
amplitude-damping and depolarizing channels respectively.

\begin{figure}[h]
        \centerline{
           \includegraphics[scale=0.9]{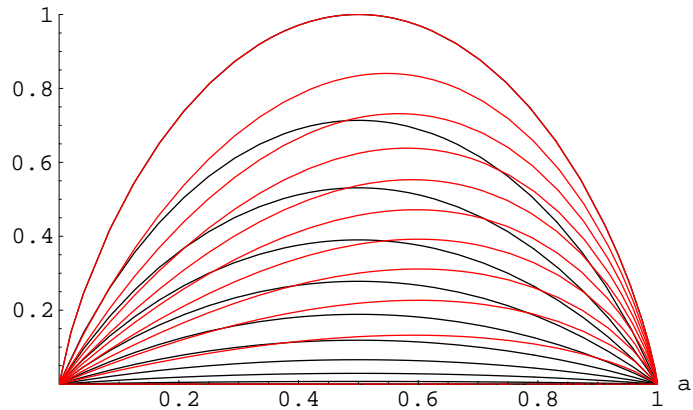}}
          \vspace*{4pt}
            \caption{The Holevo $\chi$ quantity for the amplitude
            damping channel and the depolarizing channel plotted
            as a function of $a$ for different parameter values.
            The amplitude-damping channel is represented in bold.}
            \label{DepAmpDamp}
       \end{figure}
They are plotted in Figure \ref{DepAmpDamp} for $0 \leq \gamma ,
\lambda \leq 1$. The plot above indicates that, for certain values of
$\gamma$ and $\lambda$ the maximizer for the amplitude-damping
channel lies to the right of the intersection of $\chi_1(a)$ and
$\chi_2(a)$ for the depolarizing channel, whereas that for the
depolarizing channel lies to the left. Indeed, keeping $\lambda$
fixed, we can increase $\gamma$ until the maximum of
$\chi_{AD}(\gamma)$ lies above the graph of $\chi_{Dep}$. The two
graphs then intersect at a value of $a$ intermediate between
$\half$ and the maximizer for $\chi_{AD}$. This proves that the
maximum of the minimum of the channels is in general not equal to
the minimum of the individual channel capacities.

\vspace*{-5pt}   


\begin{thebibliography}{0}
\bibitem{footNote} The maximizer for this case has also been obtained in \cite{9}, but their proof is different.
\end{thebibliography}
\end{document}